\documentstyle[a4,11pt]{article}
\title{ The Structure of the Stoichiometric and \\
Reduced SnO$_{2}$ (110) Surface}
\author{I. Manassidis, J. Goniakowski,
L. N. Kantorovich and M. J. Gillan \\
Physics Department, Keele University \\
Keele, Staffordshire ST5 5BG, U.K. \\
Corresponding author: M. J. Gillan, e-mail: pha71@keele.ac.uk}

\begin{document}
\maketitle
\begin{abstract}
First-principles calculations based on density functional theory (DFT)
and the pseudopotential method have been used to study the
stoichiometric and reduced SnO$_{2}$ (110) surface. The ionic relaxations
are found to be moderate for both the stoichiometric and
reduced surfaces, and are very similar to those found in recent
DFT-pseudopotential work on TiO$_{2}$. Removal of neutral oxygen
leaves two electrons per oxygen on the surface, which are distributed
in channels passing through bridging oxygen sites. The associated
electron density can be attributed to reduction of tin from Sn$^{4+}$
to Sn$^{2+}$, but only if the charge distribution on Sn$^{2+}$ is
recognized to be highly asymmetric. Reduction of the surface gives
rise to a broad distribution of gap states, in qualitative agreement
with spectroscopic measurements.
\end{abstract}

\section{Introduction}
Stannic oxide (SnO$_{2}$) is a wide-gap semiconductor ($E_{g}$ = 3.6 eV
\cite{agekyan}) which is used in
gas-sensing devices \cite{mcaleer},
and is also of interest as an oxidation
catalyst \cite{brown}. These
applications have stimulated intensive study of its surface
properties \cite{defresart,pacox,flavell,eriksen,erickson,
dfcox1,dfcox2,dfcox3,
themlin,shen,egdell,munnix,godin,mulheran},
concentrating mainly on the (110) surface (see fig. 1),
which is the most stable.
Because of the variable valence of Sn,
the material readily loses surface oxygen, and there have been a number of
studies of the atomic structure, electronic structure
and electrical properties of the (110) surface as a function
of temperature and oxygen partial pressure. The existence of a rather
complicated series of surface reconstructions \cite{defresart,
dfcox2,shen},
marked changes of
electronic structure in the band gap \cite{pacox,dfcox1,
themlin},
and strong variation of the
surface electrical conductivity with deviation
from stoichiometry \cite{dfcox1,shen}
are all well documented. In spite of this work, many questions
remain unresolved. For example, there is no experimental information
about the energetics of the surface, or about the nature of
surface relaxations. Nor is there a clear picture at present of
the electronic charge density at the surface, and the way this is
modified by removal of oxygen. The detailed interpretation of
the observed gap states also remains controversial.

The aim of the present work is to use first-principles calculations
to build up a detailed picture of the energetics, atomic structure
and electronic structure of both the stoichiometric and the reduced
SnO$_{2}$ (110) surface. We report calculated results for the
energy of the relaxed and unrelaxed surfaces, the relaxed atomic
positions, the valence electron distribution and the surface electronic
density of states.
The calculations have been performed using
density functional theory (DFT) and the pseudopotential
technique \cite{srivastava,aussois,revmodphys}.
These methods have a growing record of success in studying a wide
range of oxides, including MgO \cite{devitaprl,devitaprb},
Li$_{2}$O \cite{europhys}, Al$_{2}$O$_{3}$ \cite{corbas,corcer},
SiO$_{2}$ \cite{binggeli} and TiO$_{2}$ \cite{rama1,rama2},
and have been used to investigate the
structure and energetics of lattice defects and surfaces
of oxide materials, and the
process of molecular adsorption on their surfaces \cite{pugh}.
In a previous paper \cite{paper1}, we have
reported a detailed investigation of the SnO$_{2}$ and SnO perfect
crystals using the present methods. We showed there that the calculations
account satisfactorily for the equilibrium crystal structure,
phonon frequencies and electronic structure of the two materials,
and those investigations form the basis for the present work.

When considering the structure of oxide surfaces, it is important
to recall the very great differences that exist between different surfaces.
At one extreme, there is the almost complete lack of relaxation
effects revealed both by
experiments \cite{weltoncook,blanchard,urano}
and by ab initio calculations \cite{pugh,causa}
for the case of the MgO (001) surface; at the other, there are
the very large atomic relaxations leading to a major reduction in
surface energy found on the $\alpha$-alumina basal-plane
surface \cite{corbas,corcer}. In this
context, we shall see that SnO$_{2}$ (110) is an intermediate case,
showing only moderate deviations from bulk termination. Even the gross
disturbance associated with loss of surface oxygen will be shown to
lead to only rather minor displacements.

A particularly important aspect of the present work concerns the
spatial distribution of valence electrons at the surface. On a simple
ionic picture, removal of neutral oxygen would lead to the reduction
of tin from the Sn$^{4+}$ to the Sn$^{2+}$ state,
and spectroscopic studies
have given support to this \cite{themlin}.
Alternatively, we might regard the
removal of neutral oxygen as creating surface F-centres. On the first
view, we would expect the electrons left on the surface to be
concentrated on surface Sn ions. The second view might lead us to
expect electrons to be left on the site vacated by the removed
oxygen. The present first-principles calculations will give an
answer to this question, and will also allow us to investigate the
states occupied by these electrons.

There have been many spectroscopic measurements of the surface
electronic structure of SnO$_{2}$ (110). The experimental evidence
shows that on the stoichiometric surface there are no states in the
gap \cite{dfcox1}. However,
removal of surface oxygen by heating in high vacuum
leads to a broad distribution
of gap states \cite{dfcox1}.
The surprising feature is that these gap states appear
to extend up in energy from the valence band maximum (VBM). At moderate
temperatures of $\sim$ 600 K, there is a weak distribution of
states extending to ca. 1.6 eV above the VBM. As the temperature
is raised to $\sim$ 1075 K, and more oxygen is lost,
a continuous distribution of states spreads up to the Fermi level,
which lies just below the
conduction band minimum (CBM).
This implies that gap states extend almost across the entire gap.
This behaviour is surprising because in most materials it is
appropriate to consider the F-center states as pulled down from
the CBM by the Coulomb attraction of
conduction electrons to the anion vacancy. Indeed,
the electron states associated with F-centers in
{\em bulk} SnO$_{2}$ are
known to lie only $\sim$~0.15~eV below the CBM \cite{marley,samson}.
An instructive comparison here is with the electronic structure
of the (110) surface of TiO$_{2}$, which has the same crystal
structure as SnO$_{2}$. When TiO$_{2}$ (110) loses oxygen, gap
states appear below the CBM, as expected from the simple arguments,
the depth of these states below the CBM being $\sim$ 0.8 eV
\cite{henrich,smith,kurtz}.
An important purpose of the present work is to shed light on the
seemingly anomalous behaviour of SnO$_{2}$.
As we shall see, our calculations do show a broad
distribution of states in the gap for the reduced surface of SnO$_{2}$,
and we shall be able to identify these surface states as the states
occupied by electrons left by removed oxygen. We shall also be able
to confirm a conjecture by P. A. Cox {\em et al.} \cite{pacox}
that the energy of the surface
states is strongly influenced by the high polarizability of the
Sn$^{2+}$ ion, and we shall suggest a connection with the unusual
structure of the SnO crystal, which is also stabilized by this
polarizability, as pointed out in our recent paper \cite{paper1}.
So far as we are aware, no previous first-principles work on
SnO$_{2}$ surfaces has been reported, although there have been
two recent papers on the TiO$_{2}$
surface \cite{rama1,vogtenhuber}, and we shall
discuss the relation between that work and the work reported here.
There have been previous attempts to treat the surface properties
of SnO$_{2}$ (110) using non-self-consistent
tight-binding methods \cite{munnix,godin},
which will also be discussed later.

The rest of the present paper is organized as follows. The following
section gives a brief summary of the techniques used in the work.
Sec. 3 then presents our results for the equilibrium structure,
energy, valence charge distribution and density of states of the
stoichiometric (110) surface. Results for the reduced (110) surface
are described in sec. 4.
Finally, the significance of our results
and some pointers to future work are discussed in sec. 5.

\section{Techniques}
The calculations are based on density functional theory and
the pseudopotential approximation, with electron correlation
described by the local density approximation (LDA). These well established
and commonly used techniques have been extensively reviewed in the
literature (see e.g. \cite{srivastava,aussois,revmodphys}).
In the pseudopotential approach,
it is assumed that the core orbitals have exactly the same form
as in free atoms, and only valence electrons are represented explicitly
in the calculations. The valence-core interactions are described by
non-local pseudopotentials, which are generated by {\em ab initio}
calculations on isolated atoms. The solid-state calculations are
performed on periodically repeating cells, with occupied valence
orbitals represented by a plane-wave expansion. This expansion
includes all plane waves whose kinetic energy $E_{k} =
\hbar^{2} k^{2} / 2m$ ($k$ the wavevector, $m$ the electron mass)
satisfies $E_{k} < E_{\rm cut}$, where $E_{\rm cut}$ is a chosen
plane-wave cut-off energy. Convergence of the calculations with
respect to the size of basis set is achieved by systematic increase
of the plane-wave cut-off $E_{\rm cut}$.

The present calculations were performed partly using the sequential
code CASTEP ({\em C}ambridge {\em S}equential {\em T}otal
{\em E}nergy {\em P}ackage) \cite{revmodphys} running on
the Cray Y-MP at the Rutherford-Appleton Laboratory, and partly
with its parallel version
CETEP ({\em C}am\-bridge-{\em E}dinburgh
{\em T}otal {\em E}nergy {\em P}ackage) \cite{compphys}
on the 64-node Intel iPSC/860 machine at the Daresbury Laboratory.

The pseudopotentials we use are exactly the same as those used in our
earlier work on the SnO$_{2}$ and SnO perfect
crystals \cite{paper1}. Briefly,
the Sn pseudopotential was generated using the standard Kerker
scheme \cite{kerker}, treating the 5$s$ and 5$p$ levels as valence
states and the 4$d$ levels and all lower levels as core states.
For oxygen, we use a pseudopotential that was carefully optimized
using the technique of Lin {\em et al.} \cite{qteish} in order
to reduce the plane-wave cut-off as much as possible. Both
pseudopotentials are treated in the Kleinman-Bylander (KB)
representation \cite{kleinman}, with the $p$ wave treated as local
for both elements. The LDA exchange-correlation energy is represented
by the Ceperley-Alder formula \cite{ceperley,perdew}, and
Brillouin zone sampling is performed using the Monkhorst-Pack
scheme \cite{monkhorst}.

The pseudopotentials, and other aspects of the current techniques,
have been thoroughly tested in our earlier work on the SnO$_{2}$ and
SnO perfect crystals \cite{paper1},
where we made detailed comparisons with
experiment for the equilibrium crystal structures, the zone-center
phonon frequencies and the electronic densities of states.
We showed there that with the pseudopotentials we are using
a plane-wave cut-off of 1000 eV is needed to achieve good
convergence, and this is the cut-off used throughout the present work.

\section{The stoichiometric (110) surface}
A general view of the stoichiometric (110) surface of SnO$_{2}$ is shown in
fig. 1. The structure we are assuming here is the one deduced from a
wide range of measurements (see e.g. \cite{dfcox1,themlin,egdell}).
The terminology we shall use when discussing the surface
structure is as follows. The outermost ions, which are oxygen ions
forming ridges parallel to the $c$-axis, will be referred to as
bridging oxygens. Below these is a plane containing both tin and oxygen
ions; the oxygens in this plane will be called in-plane oxygens. The tin
ions in this plane occupy two kinds of site, one of which lies below
the ridge of bridging oxygens and will be call the bridging tin site. The
second kind of tin site is five-fold coordinated on the stoichiometric surface,
and we call it the five-fold tin site. Finally, we shall need to refer
to the oxygens below this plane, lying vertically beneath bridging
oxygens; we call these sub-bridging oxygens.

\subsection{Tests of vacuum width}
As in our previous work on oxide
surfaces \cite{corbas,corcer,pugh}, the calculations
are done in slab geometry. The use of this geometry is an
artifact which we are obliged to use because our electronic
structure techniques are based on periodic boundary conditions.
Our calculations are performed on a collection of ions in a box
having the shape of a parallelepiped, which is periodically repeated
in all three directions. The repeating box contains a single unit
cell of a slab of ions, and repetition in the plane of the slab has the
effect of creating an infinite slab.  Repetition in the third direction
creates an infinite stack of slabs, each slab being separated
from its neighbours by a vacuum layer.

Our real interest is, of course, in the properties of a single surface
of a semi-infinite crystal. In order to be sure that these properties
can be obtained by studying repeated slabs, one has to choose the
width of the vacuum between slabs, and the thickness of the slabs
themselves large enough so that the surfaces have only a negligible
effect on each other.
Slabs of stoichiometric SnO$_{2}$ having (110) surfaces consist of layers
containing Sn and O ions (referred to as Sn-O layers), and layers
containing only O ions. It is convenient to specify the thickness
of such a slab by giving the number of Sn-O layers. The thinnest
slab for which we can expect meaningful results consists of two
Sn-O layers. We report results both for this
system, whose slab unit cell contains 12 ions, and the four-layer
system having a unit cell of 24 ions.

We have investigated the effect of vacuum width by performing a series
of calculations on the two-layer system. For the purpose of these
tests, the ions in each slab are at their unrelaxed perfect-crystal
positions. Brillouin zone sampling is performed using the lowest-order
Monkhorst-Pack scheme \cite{monkhorst}, so that the sampling wavevectors
are ($\pm \frac{1}{4}$, $\pm \frac{1}{4}$, $\pm \frac{1}{4}$) in units
of the primitive reciprocal lattice vectors of the repeating cell.
We specify the vacuum width $L^{\prime}$ by the distance between
neighbouring slabs measured along the surface normal, defined
so that for zero width the stack of slabs becomes identical to the
infinite perfect crystal.

For these tests, we wish to know how large the vacuum width has to be
for the surface energy $\sigma$ to be independent of this width.
Table 1 shows our calculated values of the unrelaxed surface energy
of the two-layer system as a function of $L^{\prime}$. The surface
energy $\sigma$ is obtained by subtracting the energy per cell for the
case $L^{\prime} = 0$ from the energy per cell for the $L^{\prime}$ of
interest and dividing by the total surface area per cell.
Judging by the dependence of $\sigma$ on $L^{\prime}$, one sees that slab
interactions across the vacuum width are completely negligible
for $L^{\prime} \geq$ 3.28 \AA, and this is the value of $L^{\prime}$
we have adopted for all subsequent calculations. The effects of
slab {\em thickness} will become clear as we proceed.

It is useful to note that for thick vacuum layers the energy dispersion
of electron states propagating normal to the surface must vanish. This
means that there is nothing to be gained from $k$-point sampling normal
to the surface. The Monkhorst-Pack sampling set
($\pm \frac{1}{4}$, $\pm \frac{1}{4}$, $\pm \frac{1}{4}$) can therefore be
replaced by the set (0, $\pm \frac{1}{4}$, $\pm \frac{1}{4}$) without
making any difference. We have checked that this is indeed the case
for the vacuum width of 3.28 \AA.

\subsection{The relaxed structure and the valence charge density}
We have performed calculations of the equilibrium relaxed structure
for both the two-layer and four-layer systems.
Relaxation has been performed
using the steepest descents method, the criterion for equilibrium being
that the forces on all atoms should be less than 0.4 eV \AA$^{-1}$. This
criterion is enough to ensure that further reduction would change the surface
energy by less than 0.01 J m$^{-2}$ and the ionic positions by less
than 0.01 \AA; these changes are negligible for practical purposes.

Relaxation of the two-layer system causes the surface energy to
decrease from 1.96 to 1.35 J m$^{-2}$. For the four-layer system, the
relaxed surface energy is 1.50 J m$^{-2}$. The difference between these two
values is not entirely negligible, and it is clear that
interactions between the surfaces are significant for the two-layer
slab. So far as we are aware, there are no experimental values for
the surface energy of SnO$_{2}$. However, it is interesting to note
that the calculations of Mulheran and Harding \cite{mulheran}
based on an empirical interaction model for SnO$_{2}$ gave the
value of 1.38 J m$^{- 2}$ for the (110) surface.

Table 2 shows the displacements of the surface Sn and O ions away from
their perfect-crystal positions for both the two-layer and four-layer
slabs. For both the five-fold and bridging Sn ions
and for the bridging O ion, the
relaxations are along the surface normal, i.e. the (110) direction,
by symmetry. For the in-plane O ion, relaxations along both the (110)
and ($1 \bar{1} 0$) directions are possible. Relaxations are counted
positive if they are along
the outward normal, and for in-plane O, displacements
along ($1 \bar{1} 0$) are
positive if the ion moves away from the neighbouring bridging Sn.

Two conclusions emerge from this table. Firstly, the displacements
of surface ions are substantial, compared with the relaxations
at inert surfaces like MgO (001). Secondly, there are significant
differences between the results for the two slabs. In all cases, the
displacements are smaller for the thicker slab, and this explains
why the relaxed surface energy is slightly higher for the 4-layer
slab. The indication is that relaxations at the two surfaces are
interacting significantly with each other in the 2-layer slab,
and that this slab is not fully adequate for quantitative calculations.
However, the sign and general magnitude of the displacements are
the same for both slabs. More discussion of the displacements will
be given in sec. 5.

We show in fig. 2 the valence electron density on two (001) planes
which both pass through Sn and O sites. The remarkable feature of
these plots is the almost total absence of disturbance of the density
in the region away from the surface. We shall refer to these plots again
when we discuss the non-stoichiometric surface.

\subsection{Electronic structure}
We have calculated the electronic density of states (DOS) of the
relaxed two-layer and four-layer systems. As with our calculations
on the DOS of the perfect crystal \cite{paper1},
we have used the tetrahedron method
to perform the integration over the Brillouin zone.
(In fact, since $k$-point sampling is unnecessary
along the surface normal, we sample only over the
{\em surface} Brillouin zone, so that the tetrahedron method
reduces to the triangle method.) It should be stressed
that for systems of the size treated here the calculations are very
demanding, and the Brillouin zone sampling we have been able
to achieve is rather restricted. We have needed to limit the
sampling to only nine points in the irreducible wedge of the surface
Brillouin zone, and the energy resolution that this gives is limited.

The calculated DOS for the two sizes of slab are compared with the
bulk DOS obtained in our earlier work in fig. 3. As expected, there is
a close similarity between the bulk and slab DOS. However, there are
three important differences. Firstly, the slabs exhibit new structure
at the top of the valence band, in the form of a pair of sharp
peaks. Secondly, there is a slight reduction in the width of the gap.
The calculated gap for the bulk is 2.2 eV (experimental
value is 3.6 eV \cite{agekyan}),
while for the thin and thick slab systems we find
gaps of 1.6 and 1.5~eV
respectively. The new peaks at the VBM appear to be
mainly responsible for
the narrowing of the gap. The third difference is that in the slab
system states split off from the top of the O(2$s$) band.

We have studied the states at the top of the valence band in
some detail. Examination of the eigenvalues shows that there
are two states per repeating cell associated with each of the
sharp peaks at the VBM. To analyze the nature of these states,
we have calculated the electronic density $\rho_{i} ( {\bf r} )$
of each of these states separately. This density is defined as:
\begin{equation}
\rho_{i} ( {\bf r} ) =\sum_{\bf k} w_{\bf k} | \psi_{i {\bf k}}
( {\bf r} ) |^{2} \; ,
\end{equation}
where the sum goes over sampling vectors in the irreducible wedge,
$w_{\bf k}$ is a sampling weight, and $\psi_{i {\bf k}} ( {\bf r} )$
is the normalized wave function of state $i$ at sampling
vector ${\bf k}$.

We find that all four states are quite strongly localized on
bridging oxygens, with some weight also on sub-bridging oxygens;
the weight on in-plane atoms and elsewhere in the slab is small. One
of the states associated with each DOS peak can be regarded
as being localized at one of the surfaces of the slab. To illustrate
the localization of the states, we show in figure 4 a plot of
the sum of the densities $\rho_{i} ( {\bf r} )$ associated
with the two states responsible for the first peak at the VBM.
This shows the rather rapid decrease
of density as one passes into the slab along the surface normal through
the bridging oxgyen. Detailed examination of density contour plots
(not shown here) indicates that the states associated with the two
peaks have different symmetries. The states of lower energy have
mainly $p$-like character, with the axis along the surface normal.
The higher states are also $p$-like, but with the axis along
the $(1 \bar{1} 0)$ direction, in other words perpendicular to
the rows of bridging oxygens.

Although most of the reduction of the band gap is due
to the localized states at the VBM, there is some indication
of a small contribution to the reduction from states at the CBM.
This indication comes from examination of $\rho_{i} ( {\bf r} )$ for
the two lowest states of the conduction band, which show some localization
in the surface region.

The peak in the DOS split off from the O(2$s$) band contains two states
per cell. Plots of $\rho_{i} ( {\bf r} )$ show that these states
are strongly localized on the bridging oxygens, with some weight
on the sub-bridging oxygens. The form of $\rho_{i} ( {\bf r} )$
confirms the expected $s$-like character of these states.

\section{The reduced (110) surface}
\subsection{The relaxed structure and valence charge density}
All the experiments indicate that reduction of the (110) surface
occurs by removal of bridging oxygens. However, it is also clear
that the fraction of bridging oxygens removed depends on the conditions.
We have investigated two cases: first, the case where all
bridging oxygens are removed (referred to as fully reduced),
and second, the case where only half of them
are removed (half reduced). We have studied the fully reduced surface
using both the two-layer and four-layer slabs. To study the half-reduced
case, we have treated the situation where every other bridging oxygen
is removed, so that we have an ordered array of surface vacancies. For
this case, the surface unit cell has to be doubled along the $c$-axis,
and at present we can only handle this with the two-layer slab.
Views of the two reduced surfaces are shown in fig. 1. It is important
to stress that we create the reduced surface by removing oxygen
from one surface of the slab only; the other surface remains
stoichiometric.

The ionic relaxations for the fully relaxed surface are reported in
table 3. It is extremely remarkable that the displacements of the ions near
the surface are almost identical to those we have found
at the stoichiometric surface (see table 2). This strongly suggests
that the electrons left at the surface by the removal of oxygen
are exerting roughly the same forces as the original bridging oxygens.
In table 4, we show the ionic relaxations for the half-reduced surface.
In this case, because of the lower symmetry, the ionic displacements
can have components in directions for which they were zero by
symmetry at the fully reduced surface. Comparison of these results
with those for the 2-layer stoichiometric slab (table 2) shows that
the displacements are almost identical in the two cases. This
indicates once again that removal of bridging oxygen leads to
surprisingly small adjustments of the ionic positions.

The valence charge density of the relaxed fully-reduced four-layer
slab is shown in fig. 2 on the same two (001) planes for which we
showed the charge density of the stoichiometric surface. Comparing
the two pairs of plots, we see a very large change of density
distribution in the immediate neighborhood of the removed bridging
oxygen and the bridging tin directly below it. The extreme localization
of this disturbance is very striking. Not only is the density completely
unaltered on atomic layers immediately below the surface, but even
around the in-plane oxygen and five-fold tin sites there is no
visible change. It appears that the two electrons per bridging oxygen
left on the surface are localized in channels running along the (001)
direction just above the bridging tin sites and passing through the
bridging oxygen positions. This will be confirmed in greater detail
when we discuss the electronic structure below.

The valence distribution at the half-reduced surface is also shown in fig. 2.
The main differences with the plots for the fully reduced surface are
that the electron density is now lower near the bridging Sn site and greater
near the empty bridging oxygen site. This is what we should expect, since
the surface electrons are now more strongly confined between
the remaining bridging oxygens.

\subsection{Electronic structure}
The calculated densities of states for the two-layer and four-layer
fully reduced systems and for the two-layer half-reduced system are shown
in fig. 3. The major change from the stoichiometric system is the appearance
of broad distributions of states in the band gap. For the fully reduced
case, the states span the whole gap, but for the half-reduced system
there is a band of gap states roughly in the middle of the gap.
We also note the two sharp peaks at the VBM and the peak split off from
the O(2$s$) band that we found for the stoichiometric surface.
Since bridging oxygens have been removed only from one side of the slab,
we clearly expect these states to remain unchanged on the other
surface. This implies that there should be one state in each of
these peaks, and we have verified that this is the case.

We find that there is a single state responsible for the part
of the DOS spanning the gap. The variation of the energy of this
state with ${\bf k}$-vector is responsible for the spread in
energy. Density plots of $\rho_{i} ( {\bf r} )$ for this
state show that it is strongly localized in the surface region
near the bridging tin and oxygen sites.
Contour plots of $\rho_{i} ( {\bf r })$
on (001) planes passing through this site and through the bridging
tin site are shown in fig. 5. The plots show that there are
substantial concentrations of density in the region vertically
above the bridging tin site, on in-plane oxygens and on the
sub-bridging oxygens. Integration of the density in these features
shows that roughly 30 \% is in the region
above the tin site, 10 \% on each of the in-plane oxygens, 10 \% on
sub-bridging oxygen, and the remainder in small features elsewhere.

\section{Discussion}
In assessing our results, it is important to bear in mind the
strengths and weaknesses of density functional theory, which forms
the basis of all our calculations. DFT is a theory of the ground state,
and we expect it to give an accurate description of the energetics
of the system in the ground state, which in the present case includes
surface energies and the equilibrium positions of the atoms in
the relaxed surface. The theory is also designed to give accurate
results for the valence electron distribution, and for related
quantities such as the distribution of electrostatic potential.
On the other hand, the principles of DFT give us no right to
expect accurate results for excited-state properties. It is well
known, for example, that DFT generally underestimates band gaps.
As we reported in our previous paper \cite{paper1}, our
calculated band gap for bulk SnO$_{2}$ is 2.2 eV, which is
$\sim$ 40 \% smaller than the experimental value of 3.6 eV. One therefore
needs to be more cautious about the interpretation of excited-state
results. Nevertheless, it has been found that DFT does usually
give a reliable semiquantitative guide to electronic densities
of states, and we shall certainly draw important conclusions
from our results for such quantities.

We have shown that atomic relaxations at both the stoichiometric
and reduced (110) surfaces are rather moderate, being on the
order of 0.1 - 0.3 \AA. This is large compared with the very tiny
displacements on the MgO (001) surface \cite{pugh,weltoncook,blanchard,
urano,causa},
but small compared with the
giant displacements at the $\alpha$-Al$_{2}$O$_{3}$ basal-plane
surface \cite{corbas,corcer}. We find
that five-fold coordinated Sn moves into the
surface, while bridging Sn and in-plane O move out. The separation
between bridging O and bridging Sn decreases markedly. These
qualitative features are the same as found by Ramamoorthy
{\em et al.} \cite{rama1} in their DFT-pseudopotential
calculations on the TiO$_{2}$ (110)
surface, and the magnitudes of their displacements are also very
similar to ours. The only significant difference is that the absolute
displacement of bridging O in TiO$_{2}$ is inwards, whereas we find
it be outwards for SnO$_{2}$. Independent DFT calculations on
TiO$_2$ (110) have also been performed by Vogtenhuber {\em et al.}
\cite{vogtenhuber} using the FLAPW technique. These authors find
qualitatively similar results.

Relaxation has a significant effect on the surface energy, reducing
it from 1.96 to 1.50 J~m$^{-2}$ for the stoichiometric surface. This
relaxed surface energy is quite close to the value of 1.38 J~m$^{-2}$
calculated by Mulheran and Harding \cite{mulheran}
using an empirical interaction
model. This provides support for the realism of the ionic model
they use \cite{freeman}.
Unfortunately,
Mulheran and Harding do not report results for the surface
relaxations they obtain with the model. It is also worth noting that
our calculated surface energy is rather similar to the
value of 0.9~J~m$^{-2}$ found by Ramamoorthy {\em et al.} for the
relaxed TiO$_{2}$ (110) surface.

On both the stoichiometric and reduced surfaces, we find marked changes
of electronic structure compared with the bulk. The rather well
localized surface states on bridging oxygen that we have found on the
stoichiometric surface are not unexpected. These states, both above
the O(2$s$) band and at the VBM are localized largely on bridging
oxygen, which is in an exposed position where one would expect a reduced
Madelung potential and hence a reduction in the binding of states.
So far as we are aware, there is at present no experimental evidence
for the existence of surface O(2$s$) states. Detailed UPS measurement
have been performed on the stoichiometric SnO$_{2}$ (110) surface, and no
evidence has been found for surface states at the VBM. Because of
this rather surprising disagreement, we have tested the robustness
of this aspect of our calculations by repeating them with an
independently generated pseudopotential, but this makes no difference.
It is conceivable that the problem might arise from our treatment
of the Sn(4$d$) states as core states, or even from deficiencies
of the LDA, but it would require a substantial effort
to test either of these possibilities.

At the reduced surface, we have shown that the two electrons left
by the removal of neutral oxygen occupy surface states localized
in the region of the empty bridging oxygen sites
and the adjacent bridging tin sites. More accurately,
their density is concentrated immediately above the bridging tin
atoms, with some weight on in-plane oxygens and elsewhere.
This finding provides direct support for the common assumption
that oxygen loss leads to reduction of tin from Sn$^{4+}$ to Sn$^{2+}$.
However, our results make it clear that if one does wish
to use the Sn$^{2+}$ description, then it is crucial to recognize the
extreme distortion of this ion. The change of electron density caused by
reduction occurs entirely on the vacuum side of the bridging Sn ions,
which, if considered as Sn$^{2+}$ ions, must be regarded as maximally
polarized. A high degree of polarization of these ions is expected, because
they are very polarizable, and oxygen removal makes their invironment
very unsymmetrical. Their high polarizability arises from the rather
small energy difference between 5$s$ and 5$p$ states and the large
spatial overlap of these states. As emphasized in our earlier
paper \cite{paper1}, the Sn$^{2+}$ ion in the SnO perfect crystal is also
in a very unsymmetrical environment, and we showed that there also the
Sn ion exhibits maximal polarization. It would be interesting to know the
valence charge distribution at the reduced TiO$_{2}$ surface,
but this appears not to have been examined in
the first-principles calculations of
Ramamoorthy {et al.} \cite{rama1,rama2}.
The lowest-lying conduction-band states in that case have Ti(3$d$)
character, and the weak spatial overlap between 3$d$ and 4$p$ states
will make the reduced Ti ion much less polarizable. The surface
charge distribution of TiO$_{2}$ should therefore be rather different
from that of reduced SnO$_{2}$.

On the fully reduced surface, we find a broad band of surface states
spanning the entire gap, and we have demonstrated that occupation of
these states is responsible for the surface electron density
above bridging Sn sites.
On the half-reduced surface,
the band of gap states contracts to have a width of $\sim$~1.5~eV, and lies
$\sim$ 1.1 eV above the VBM. In comparing with experiment, it
must be remembered that the real surface is likely to be significantly
disordered. Experimentally \cite{dfcox1},
a weak distribution of states near the
bottom of the gap is observed for small degrees of reduction. As
further oxygen is removed, the distribution spreads upwards to the
Fermi level, which is just below the top of the gap. It seems clear
that these must be essentially the same as the gap states we find.

The large width of the band of gap states is not surprising, since
the electrons in these states are able to propagate almost freely
along the $c$-axis. The width of $\sim$~1.5~eV we find
for the half-reduced surface is also reasonable. It is interesting
to note that recent calculations on periodic arrays of F-centers
in bulk MgO \cite{wang} find a band
width of gap states of 1.6~eV when the
spacing between F-centers is 5.9~\AA, and the F-center states
have been found to have a substantial amplitude on neighboring
oxygens. The energy of the gap state is also understandable, following
an argument originally due to P. A. Cox {\em et al.} \cite{pacox}.
We have seen that the
Sn$^{2+}$ ions are highly polarized. This must arise from a strong
mixing of the 5$s$ and 5$p$ states caused by surface fields, and this
mixing will displace one of the hybrid states downwards in energy.
Alternatively, we can adopt the F-center viewpoint and note that
F-center levels are commonly found rather deep in the gap. For
example, in bulk MgO the F-center levels are roughly 5~eV below the
CBM. In very recent DFT calculations, we have found that surface
F-centers in MgO have their levels at almost the same energy as
as those in the bulk \cite{lev}.
This still leaves an unresolved question, however, concerning
the relation between bulk and surface F-centers in SnO$_{2}$.
It is rather well established that the donor states involved
in $n$-type SnO$_{2}$ are oxygen vacancies, and that the
binding energy of electrons to these vacancies is no more than
$\sim$~0.15 eV \cite{marley,samson}.
This implies that bulk and surface F-centers
in SnO$_{2}$ must have very different properties, and the
reasons for this need to be explored. We hope to return to this
question in the future.

The insights we have gained allow us to comment on the predictions
of tight-binding calculations. The recent tight-binding work
of Godin and LaFemina \cite{godin} made a thorough investigation
of the stoichiometric and fully reduced  SnO$_{2}$ (110)
surfaces. No states were found in the gap for either surface. This
was consistent with previous tight-binding work of Munnix and
Schmeits \cite{munnix}, who studied oxygen vacancies on the
SnO$_{2}$ (110) surface and found no gap states. These results
are in conflict with what we have found.
It is also
noteworthy that tight-binding calculations on the TiO$_{2}$ (110)
surface \cite{munnixtio2}
failed to find any gap states associated with oxygen
vacancies, even though the existence of such states has been
well established for many years \cite{henrich,smith,kurtz}.
In our view, the
tight-binding models used in the above work lack an
essential piece of physics.
We have argued from our results that if one adopts an LCAO
viewpoint then Sn$^{2+}$ states
are crucially influenced by cation $s$-$p$ mixing caused by
surface fields. But such a mechanism is completely lacking from the
tight-binding models used in the above work.
There is no term in the Hamiltonians
which accounts for this mechanism of
$s$-$p$ mixing, and no inclusion of any kind
of effect arising from surface fields. Very recently,
self-consistent tight-binding calculations on TiO$_{2}$ (110) and other
oxide surfaces have been reported, in which Coulombic effects are
approximately included \cite{goniakowski,halley}.
Although only the stoichiometric surface
has been studied so far by these methods, it is already clear from
these papers that self-consistency has a major effect on the surface
electronic states.

Finally, we note that much of the practical interest in SnO$_{2}$
surfaces centers on the adsorption and reactions of molecules.
We are currently using the results of the present work as a basis
for investigating the dissociation of H$_{2}$O at the SnO$_{2}$
(110) surface \cite{gilgon}, and we hope to report on this soon.

\section*{Acknowledgments}
I.M. thanks the SERC for a research studentship and AEA Technology for
partial financial support. The calculations were performed partly
on the Intel iPSC/860 machine at Daresbury Laboratory and
partly on the Cray Y-MP at the Rutherford-Appleton Laboratory under
SERC grant GR/J59098. Subsidiary analysis employed local computer
equipment provided by SERC grants GR/H31783 and GR/J36266. Useful
discussions with A. H. Harker, G. Thornton and P. A. Cox are
gratefully acknowledged. We are also indebted to J. Holender,
who has been an unfailing source of help during this work.

\newpage

\begin{table}[p]
$$
\begin{array}{cc}
\mbox{vacuum width (\AA)} &
\mbox{surface energy }(J\cdot m^{-2})
\vspace{0.2cm} \\
\hline
1.64&1.68\\
3.28&1.96\\
4.92&2.00\\
6.56&2.00\\
\hline
\end{array}
$$
\caption[Surface energy for the unrelaxed surface
for different vacuum thickness.]
{Calculated energy per unit area of the unrelaxed (110)
stoichiometric surface of SnO$_{2}$ for different vacuum
widths.}
\end{table}

\begin{table}[p]
$$
\begin{array}{lccc}
\mbox{atoms}&\mbox{no. of layers}&
\mbox{$\Delta x$ (\AA)}&\mbox{$\Delta y$ (\AA)} \\
\hline
\mbox{bridging tins}    &2&+0.26&-\\
\mbox{bridging tins}    &4&+0.15&-\\
\mbox{5-fold tins}      &2&-0.21&-\\
\mbox{5-fold tins}      &4&-0.15&-\\
\mbox{bridging oxygens} &2&+0.11&-\\
\mbox{bridging oxygens} &4&+0.02&-\\
\mbox{in-plane oxygens}  &2&+0.11&+0.05\\
\mbox{in-plane oxygens}  &4&+0.07&+0.05\\
\hline
\end{array}
$$
\caption{Relaxed positions of ions at the stoichiometric surface.
Calculated values are given for the displacements of the ions from
their perfect-lattice positions for the two-layer and four-layer slabs.
Cartesian components of the displacements are given along the surface
normal (110) direction ($\Delta x$) and along the (1$\bar{1}$1)
direction ($\Delta y$). Dashes indicate values that are zero by
symmetry. The nomenclature for surface ions is explained in the text.}
\end{table}

\begin{table}[p]
$$
\begin{array}{lccc}
\mbox{atoms}&\mbox{no of layers}&
\mbox{$\Delta x$ (\AA)}&\mbox{$\Delta y$ (\AA)} \\
\hline
\mbox{bridging tins}  &2&+0.24&-\\
\mbox{bridging tins}  &4&+0.23&-\\
\mbox{5-fold tins}    &2&-0.17&-\\
\mbox{5-fold tins}    &4&-0.18&-\\
\mbox{in-plane oxygens}&2&+0.09&+0.07\\
\mbox{in-plane oxygens}&4&+0.08&+0.07\\
\hline
\end{array}
$$
\caption{Relaxed positions of ions at the fully reduced surface.
Cartesian components of displacements away from perfect-lattice
positions are given, following the notation and nomenclature
of table 2.}
\end{table}

\begin{table}[p]
$$
\begin{array}{lccc}
\mbox{atoms}&
\mbox{$\Delta x$ (\AA)}&\mbox{$\Delta y$ (\AA)}&\mbox{$\Delta z$ (\AA)}\\
\hline
\mbox{bridging tins}&+0.27&-&+0.03\\
\mbox{5-fold tins (1)}&-0.23&-&-\\
\mbox{5-fold tins (2)}&-0.24&-&-\\
\mbox{bridging oxygens}&+0.13&-&-\\
\mbox{in-plane oxygens}&+0.11&+0.04& 0.00 \\
\hline
\end{array}
$$
\caption{Relaxed positions of ions at the half-reduced surface. Cartesian
components of displacements away from perfect-lattice positions
are given, following the notation and nomenclature of table 2. The
displacement component denoted by $\Delta z$ is along the (001)
direction.}
\end{table}

\newpage
\section*{Figure Captions}
Fig. 1: Perspective view of the stoichiometric (upper panel),
fully reduced (middle) and
half-reduced (lower) surfaces of SnO$_{2}$. Small dark spheres and
large light spheres represent Sn and O respectively.

\noindent
Fig. 2: Contour plots of valence electron density (units: 10$^{-2}$
electron/\AA$^{3}$) on (001) planes passing through the slabs used
for studying the stoichiometric (panels a and b), fully reduced
(c and d) and half-reduced (e and f) (110) surfaces. For each surface,
two planes are shown, passing through the in-plane oxygen and bridging
tin sites (a, c, e) and the bridging oxygen and fivefold tin sites
(b, d, f). Regular Sn sites are marked by the symbol $\times$.

\noindent
Fig 3: Electronic DOS for the perfect crystal (a),
the two-layer stoichiometric
(b) and four-layer stoichiometric (c) slabs, the two-layer fully
reduced (d) and four-layer fully reduced (e) slabs, and the two-layer
half-reduced slab (f). In all cases, the dashed line marks the
energy of the highest occupied state.

\noindent
Fig. 4: Plot of electronic density associated with the two states in
the first peak at the top of the valence band in the four-layer
stoichiometric slab. The line on which the density is plotted passes
through the bridging and sub-bridging oxygen sites, which are marked
by arrows.

\noindent
Fig. 5: Contour plots of electron density (units: electrons/\AA$^{3}$)
associated with the single state (per removed oxygen) spanning the
band-gap for the four-layer fully reduced slab. Plots are on
(001) planes passing through the bridging tin and in-plane oxygen
sites (panel a) and bridging and sub-bridging oxygen sites (b).
Intersections of the guide lines mark the significant sites: in
panel (a) the intersection below the centre is the bridging
tin, and the intersections on either side are in-plane oxygens;
in panel (b), the intersection at the centre is the bridging oxygen site and
the lower intersection is sub-bridging oxygen. Distances are
in \AA.

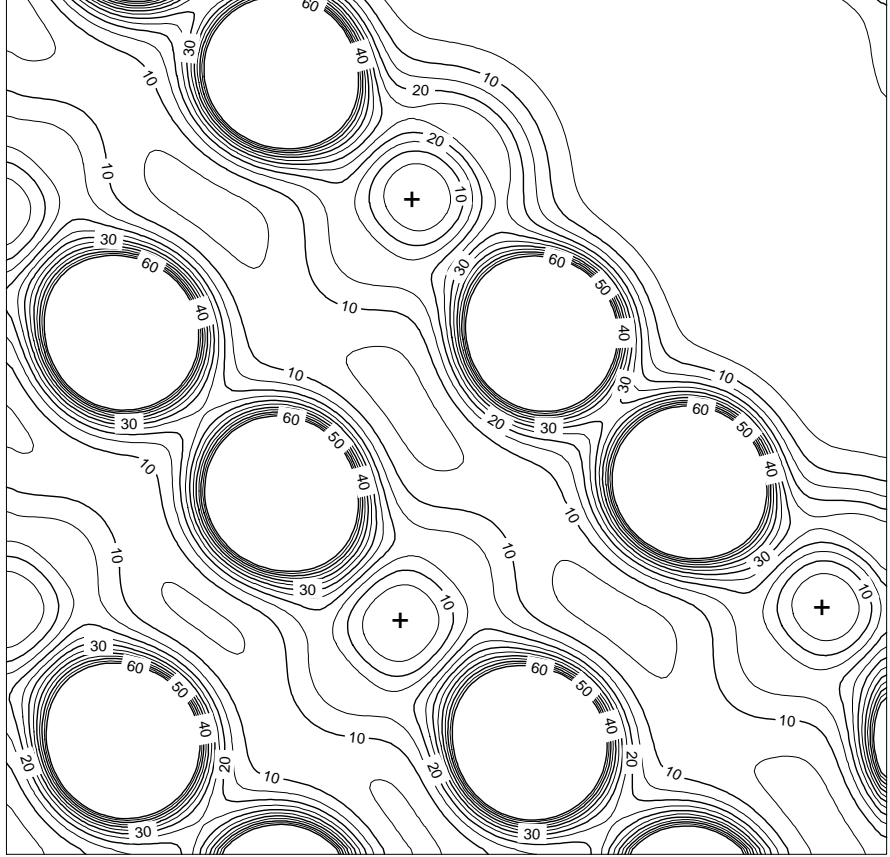
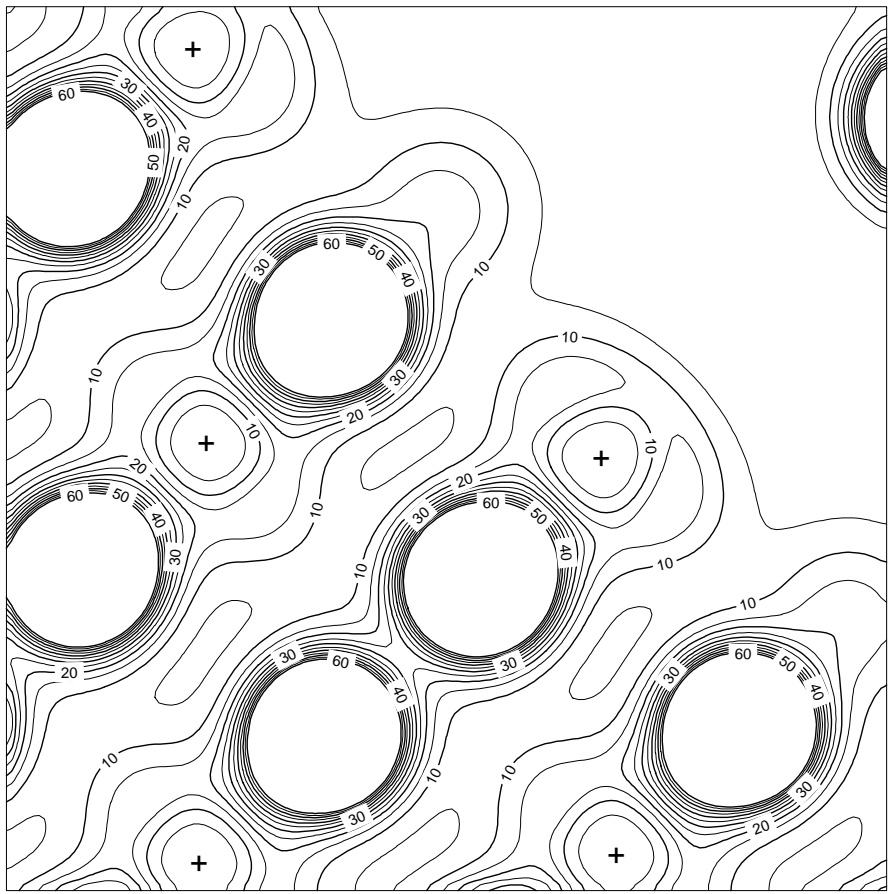
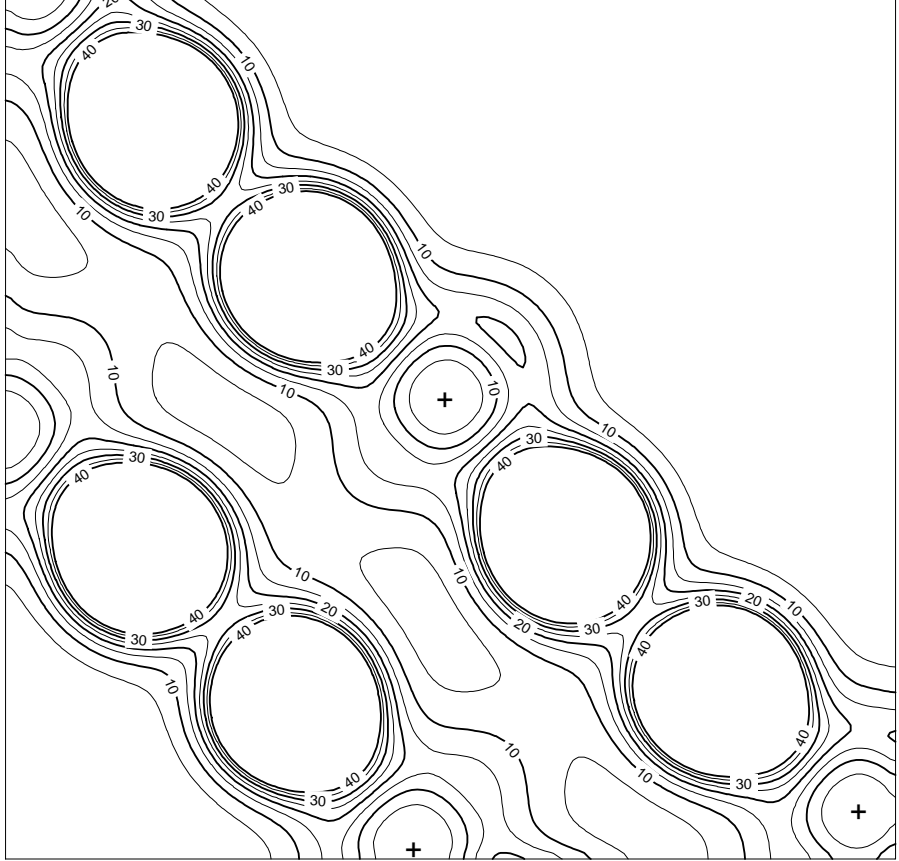
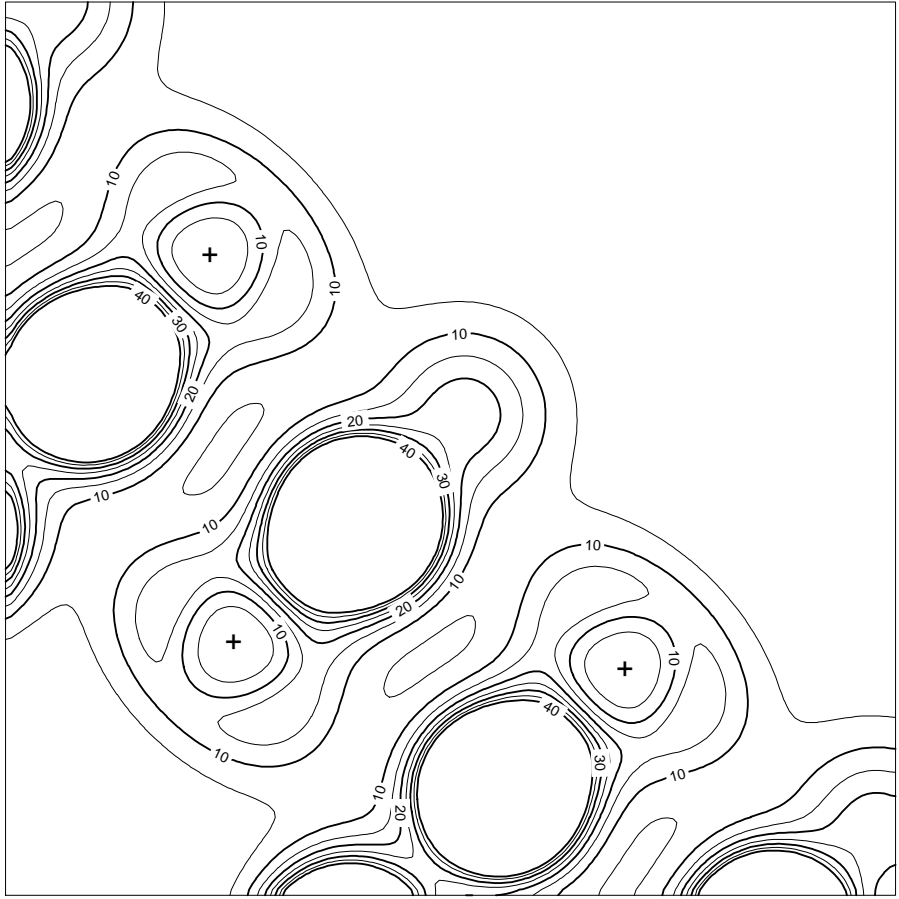
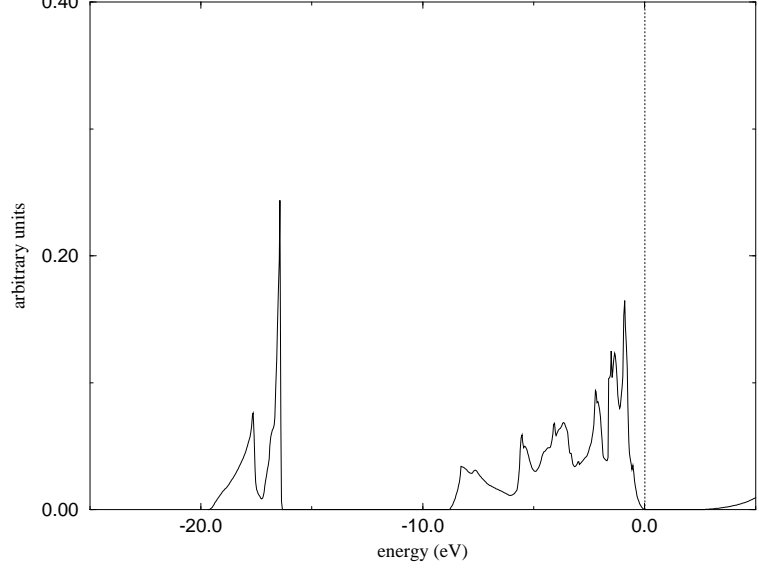
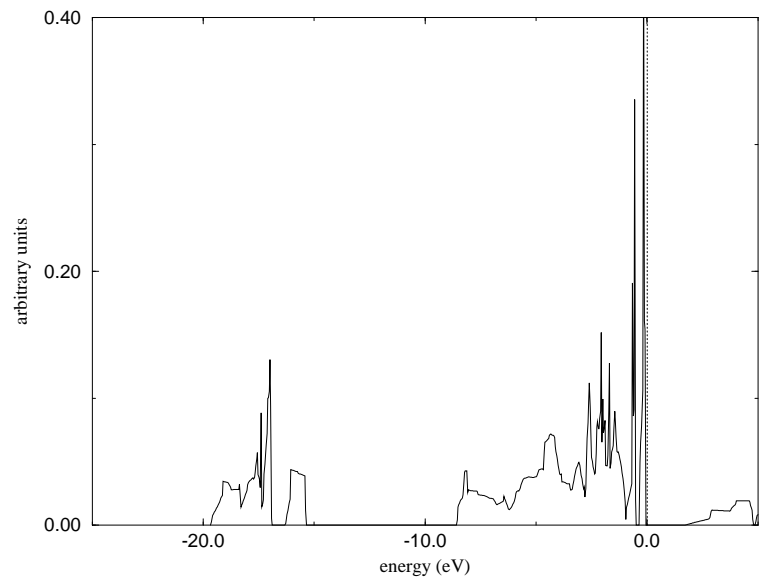
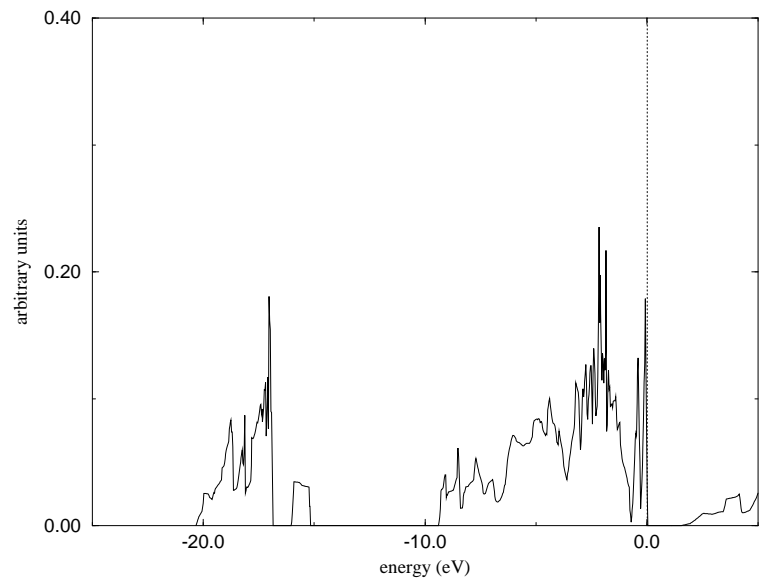
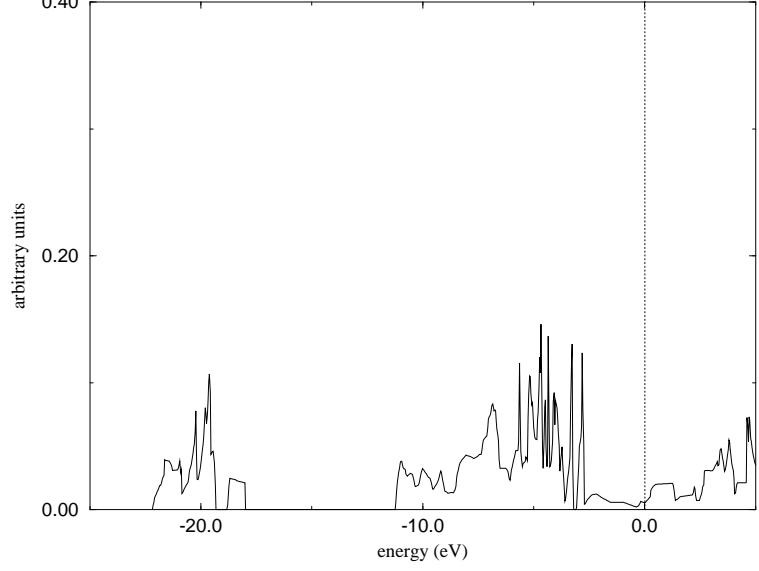
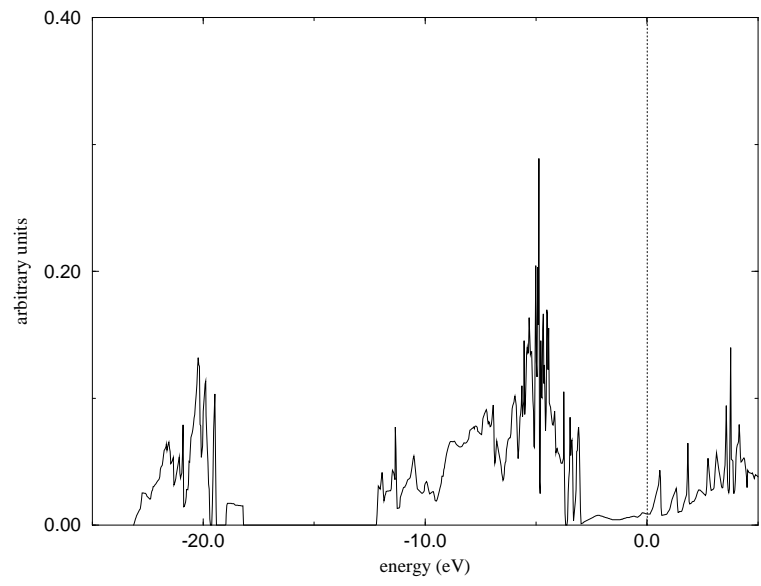
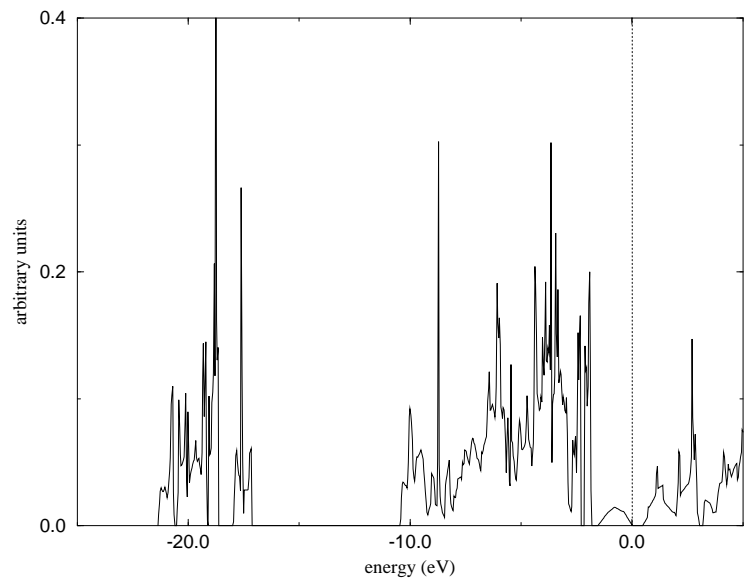
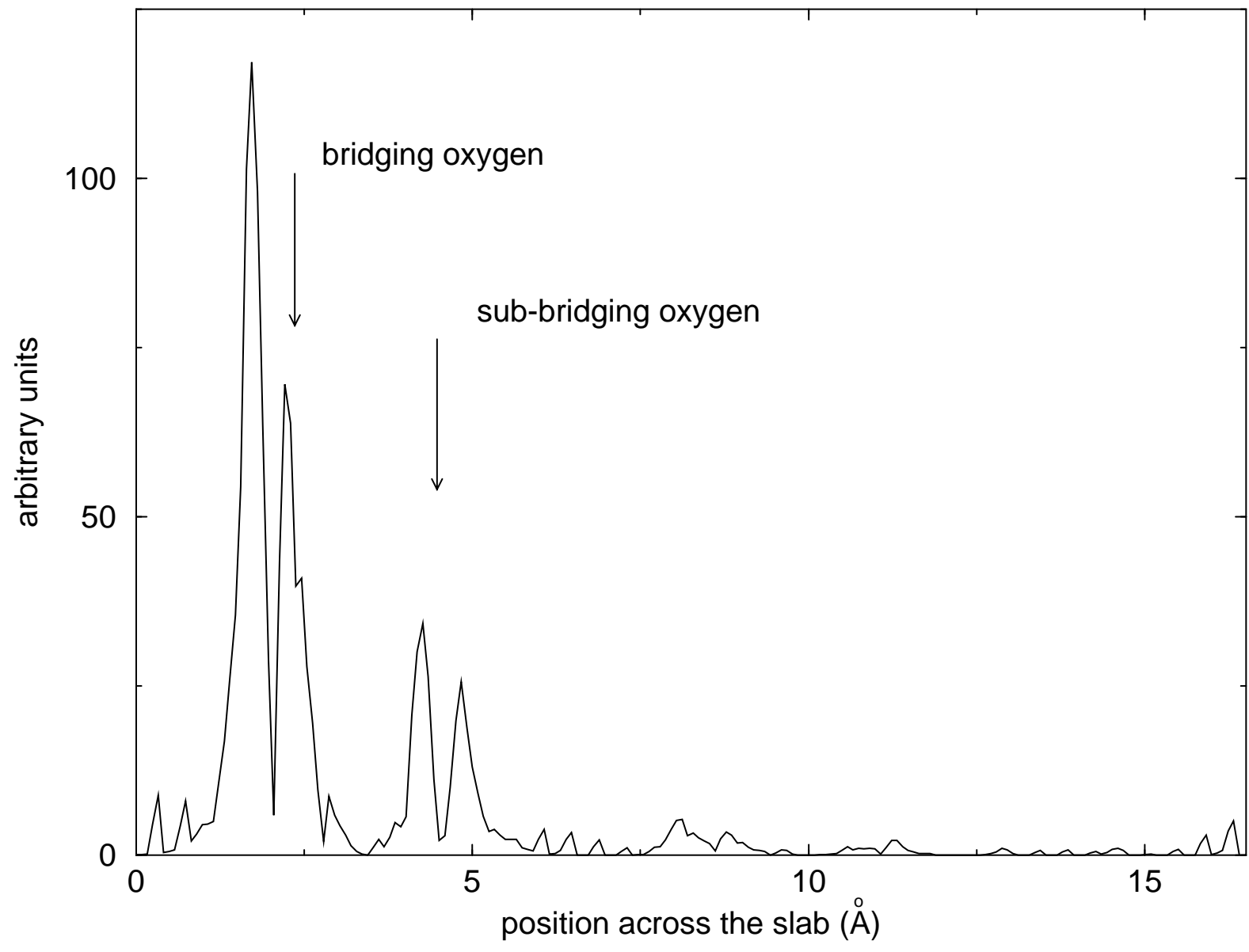
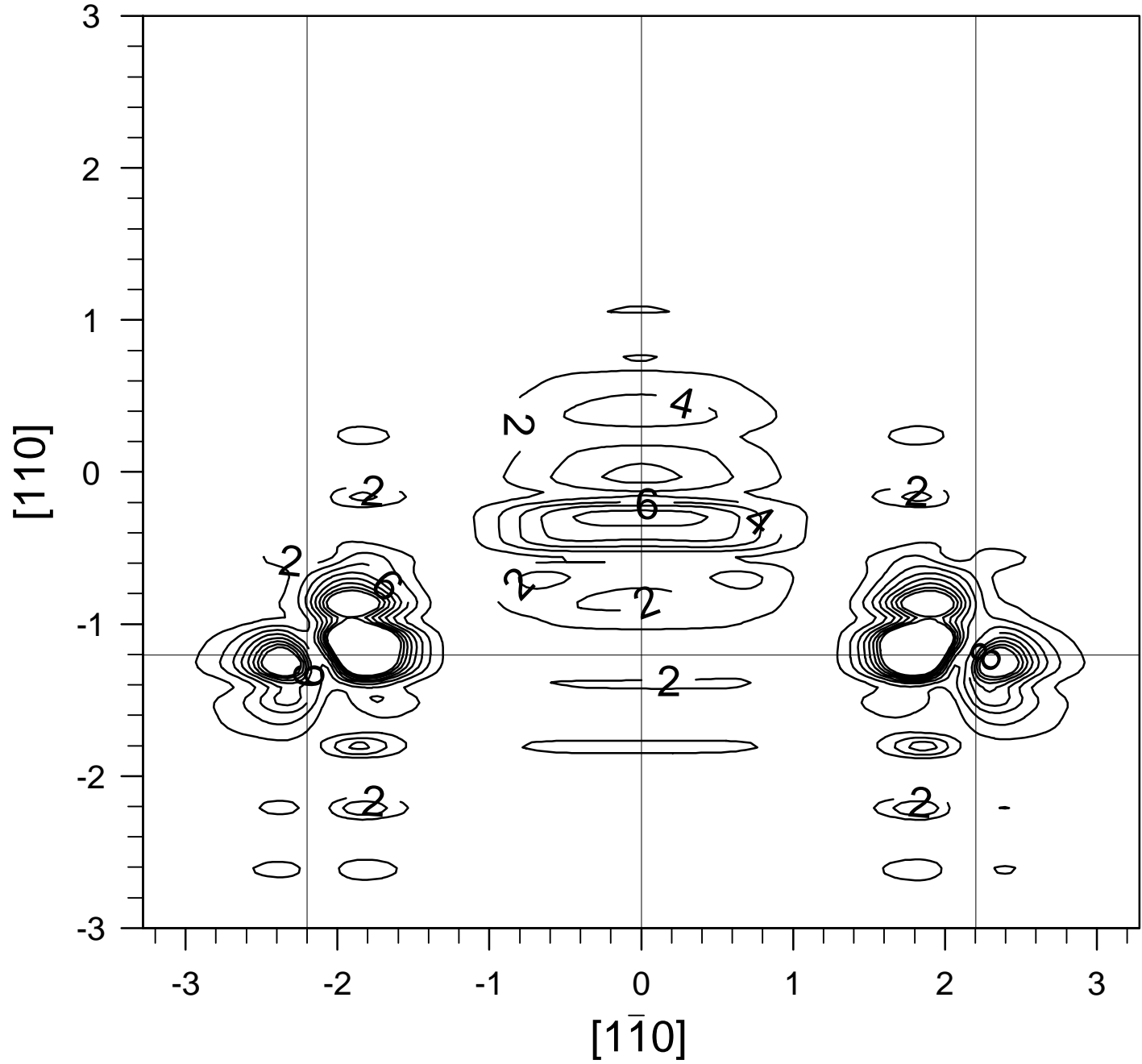
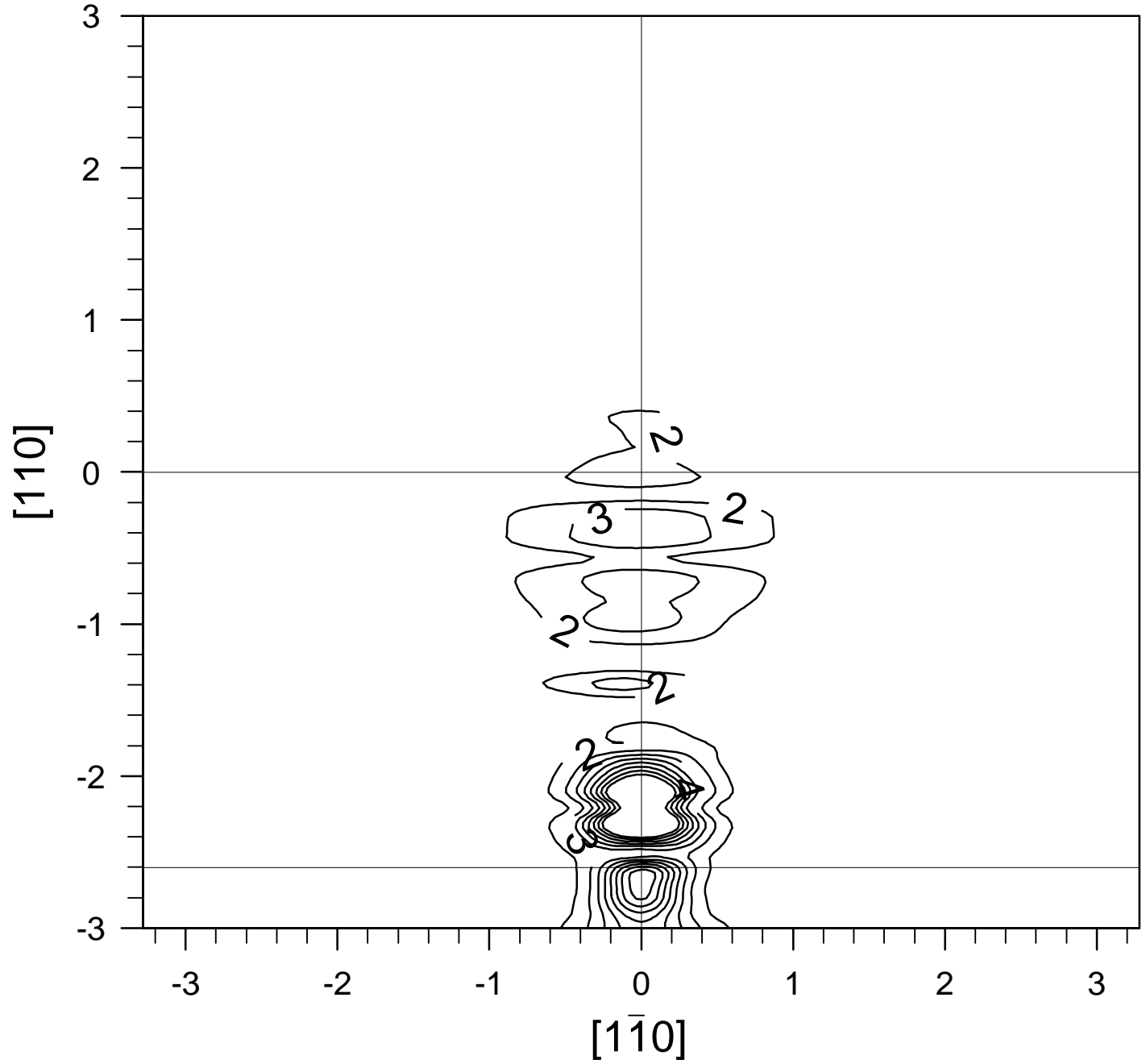
\end{document}